\documentclass[conference]{IEEEtran}
\IEEEoverridecommandlockouts

\usepackage{cite}
\usepackage{amsmath,amssymb,amsfonts}
\usepackage{algorithmic}
\usepackage{hyperref}
\usepackage{graphicx}
\usepackage{textcomp}
\usepackage{xcolor}
\usepackage{float}
\usepackage{subcaption}
\usepackage{slashbox}

\usepackage{multirow}
\usepackage{graphicx}

\def\BibTeX{{\rm B\kern-.05em{\sc i\kern-.025em b}\kern-.08em
    T\kern-.1667em\lower.7ex\hbox{E}\kern-.125emX}}
\begin{document}

\title{CinC-GAN for Effective $F_0$ prediction for Whisper-to-Normal Speech Conversion
}

\author{\textit{Maitreya Patel, Mirali Purohit, Jui Shah, and Hemant A. Patil} \\
Speech Research Lab, DA-IICT, Gandhinagar-382007, India. \\
E-mail: \{maitreya\_patel, purohit\_mirali, jui\_shah, hemant\_patil\}@daiict.ac.in}

\maketitle

\begin{abstract}
Recently, Generative Adversarial Networks (GAN)-based methods have shown remarkable performance for the Voice Conversion and WHiSPer-to-normal SPeeCH (WHSP2SPCH) conversion. One of the key challenges in WHSP2SPCH conversion is the prediction of fundamental frequency ($F_0$). Recently, authors have proposed state-of-the-art method Cycle-Consistent Generative Adversarial Networks (CycleGAN) for WHSP2SPCH conversion. The CycleGAN-based method uses two different models, one for Mel Cepstral Coefficients (MCC) mapping, and another for $F_0$ prediction, where $F_0$ is highly dependent on the pre-trained model of MCC mapping. This leads to additional non-linear noise in predicted $F_0$. To suppress this noise, we propose Cycle-in-Cycle GAN (i.e., CinC-GAN). It is specially designed to increase the effectiveness in $F_0$ prediction without losing the accuracy of MCC mapping. We evaluated the proposed method on a non-parallel setting and analyzed on speaker-specific, and gender-specific tasks. The objective and subjective tests show that CinC-GAN significantly outperforms the CycleGAN. In addition, we analyze the CycleGAN and CinC-GAN for unseen speakers and the results show the clear superiority of CinC-GAN.
\end{abstract}

\begin{IEEEkeywords}
Whisper-to-Normal Speech, Non-parallel, $F_0$ prediction, CycleGAN, CinC-GAN.
\end{IEEEkeywords}

\section{Introduction}
\label{sec:intro}

Whisper and normal speech are different way of communication. People generally use normal mode of speech in regular life, however in some cases, people need to keep their conversation private such as, during phone calls in public places, in meeting, library, hospital etc, where people adopt to use whisper mode conversation \cite{hansen2018whisper}. Whisper and normal speech are cross-domain entities, as it differs in terms of speech production and perception \cite{hansen2018whisper, nirmeshmlslp, illa2017comparative}. Given a speech, whether it is normal or not is depend on arrangements of larynx, and particularly on glottis \cite{wallis2004vocal, mattiske1998vocal, sulica2013vocal, rubin2007vocal}. Sometimes because of accident or disease, people are not able to produce normal speech, because the parts which take part in speech production get affected. Also losing the normal way of speaking will significantly affect the person's life. When people speak in normal style, vocal folds vibrates with some specific fundamental frequency (i.e., $F_0$) while this is not the case in whisper speech \cite{hansen2018whisper, quatieri2006discrete}. In addition, current speech processing systems do not perform efficiently on any kind of speech except on normal speech. Therefore, WHSP2SPCH conversion task is necessary.

One of the challenging problem in WHSP2SPCH conversion is $F_0$ prediction. However, $F_0$ is encapsulated in an intricate way in the whispered speech. The presence and absence of $F_0$ is the key difference between normal \textit{vs.} whispered speech \cite{konno2016whisper, meyer1956realization, itoh2001acoustic}. At the acoustic-level, there is difference between voiced and unvoiced speech, and statistical voice conversion (VC)-based methods are able to do such conversion \cite{toda2012statistical}. Attempts have been made in the literature for VC, such as GMM, Conditional Variational AutoEncoders (CVAE), CycleGAN-VC, etc. \cite{stylianou1998continuous, kingma2013auto, chen2014voice, kaneko2018cyclegan, adaganvc2019, Kaneko2019cycle}. For WHSP2SPCH conversion attempts have been made in the literature using parallel data only. Such as LSTM, MSpeC-Net, DiscoGAN, CycleGAN, etc. are proposed in the literature \cite{malaviya2020mspec, nirmeshmlslp, konno2016whisper, meenakshi2018whispered, janke2014fundamental, mcloughlin2013reconstruction, mcloughlin2015reconstruction, toda2012statistical, tran2009multimodal}. Moreover, CycleGAN has shown state-of-the-art result for WHSP2SPCH conversion including $F_0$ prediction on parallel data, which relies on the availability of particular speaker's whisper, and normal speech \cite{mihir2019EUSIPCO}. However, this is not feasible and it is impractical too. Moreover, parallel data requires time-alignment as pre-processing. In addition, traditional method uses 2-step sequential method for WHSP2SPCH conversion \cite{mihir2019EUSIPCO, Patel2019}. For CycleGAN based conversion, in first step, one CycleGAN is trained for cepstral feature mapping of whisper to normal speech, and in second step, another CycleGAN is trained for $F_0$ prediction, which heavily relies on previously trained CycleGAN \cite{mihir2019EUSIPCO}. Because of the imperfect cepstral feature mapping, noise is introduced in the output. Due to the non-linear DNN layers, it is non-linear noise. Therefore, significant non-linear noise is added in $F_0$ prediction. 

Although CycleGAN gives the state-of-the-art result, there is still a gap between the original and converted normal speech in terms of naturalness \cite{mihir2019EUSIPCO}. To reduce this gap and overcome above limitations, we propose CinC-GAN for non-parallel WHSP2SPCH conversion task, including $F_0$ prediction in non-parallel mode. CinC-GAN is designed specifically for effective $F_0$ prediction, which is important factor for naturalness. Here, CinC-GAN uses joint training methodology, where acoustic mapping, and $F_0$ prediction is done simultaneously. The objective result shows that CinC-GAN is able to suppress the non-linear noise in $F_0$ prediction. Therefore, $F_0$-$RMSE$ is decreased by 29.8\% and 82.2\% compared to the baseline for speaker and gender-specific tasks, respectively. Subjective evaluation shows that CinC-GAN helps to bring the converted normal speech more closer to the original normal speech compared to the baseline (CycleGAN). In objective and subjective evaluations, gender-specific task contains analysis on seen and unseen speakers. In addition, CinC-GAN maintains the naturalness for gender-specific task (for seen and unseen speakers), whereas CycleGAN degrades its result and produces whisper speech.

\section{Conventional Cycle-GAN}
\label{sec:cyclegan}

Let $x \epsilon R^N$ and $y \epsilon R^N$ be the cepstral features of whisper (X) and normal (Y) speech, respectively, where $N$ is the dimension of a feature vector. In CycleGAN, two generators are used, $G_{X \rightarrow Y}$  and $G_{Y \rightarrow X}$, where $G_{X \rightarrow Y}$ maps the cepstral features of $X$ to $Y$, whereas mapping $G_{Y \rightarrow X}$ does the opposite (i.e., $Y$ to $X$). In addition, we have two discriminators $D_X$ and $D_Y$, whose role is to predict whether its input is from the distribution $X$ and $Y$ or not, respectively.

\begin{figure}[!h]
    \centering
    \includegraphics[width=0.5\textwidth]{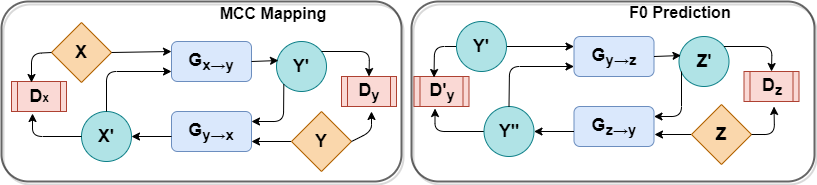}
    \caption{Conventional CycleGAN. After \cite{zhu2017unpaired}.}
    \label{fig:cyclegan}
\end{figure}

In CycleGAN, there are three types of losses, cycle-consistent loss, adversarial loss, and identity loss, as described below.

\textbf{Adversarial loss:} To make converted normal speech indistinguishable from the original, we use adversarial loss. Here, we use least square error loss instead of traditional binary cross-entropy loss, which is defined as: 

\begin{equation}
\begin{split}
\mathcal{L}_{adv}(G_{X \rightarrow Y},& D_Y) =  \mathbb{E}_{y \sim P_{Y}(y)}[(D_Y(y) - 1)^2] \\ &+ \mathbb{E}_{x \sim P_X(x)}[(D_Y(G_{X \rightarrow Y}(x)))^2].
\end{split}
\label{eq:cinc_adv}
\end{equation}

\textbf{Cycle-consistent loss:} The main idea behind this loss is to map the distribution between original and reconstructed data. In addition, this loss tries to preserve contextual information across different speech. This loss allows us to do non-parallel WHSP2SPCH conversion. The loss is defined as:

\begin{equation}
\begin{split}
&\mathcal{L}_{cyc}(G_{X \rightarrow Y}, G_{Y \rightarrow X}) \\ &=  \mathbb{E}_{x \sim P_X(x)} [\| G_{Y \rightarrow X}(G_{X \rightarrow Y}(x)) - x\|_1 ] \\ &+ \mathbb{E}_{y \sim P_Y(y)} [\| G_{X \rightarrow Y}(G_{Y \rightarrow X}(y)) - y\|_1].
\end{split}
\label{eq:cinc_cyc}
\end{equation}

\textbf{Identity-mapping loss:} To encourage preservation of input linguistic content (as suggested in \cite{zhu2017unpaired}), identity loss is used:

\begin{equation}
\begin{split}
\mathcal{L}_{id}(G_{X \rightarrow Y}, G_{Y \rightarrow X}) &=  \mathbb{E}_{x \sim P_X(x)} [\| G_{Y \rightarrow X}(x) - x\|_1 ] \\ &+ \mathbb{E}_{y \sim P_Y(y)} [\| G_{X \rightarrow Y}(y) - y\|_1].
\end{split}
\label{eq:cinc_idnt}
\end{equation}

The total loss function is defined as:

\begin{equation}
\begin{split}
&\mathcal{L}_{full} =  \mathcal{L}_{adv}(G_{X \rightarrow Y}, D_Y) + \mathcal{L}_{adv}(G_{Y \rightarrow X}, D_X)\\ &+ \lambda_{cyc}\mathcal{L}_{cyc}(G_{X \rightarrow Y}, G_{Y \rightarrow X}) + \lambda_{id}\mathcal{L}_{id}(G_{X \rightarrow Y}, G_{Y \rightarrow X}).
\end{split}
\label{eq:cinc_full}
\end{equation}

Where the values of $\lambda_{cyc}$ and $\lambda_{id}$ are 10 and 5, respectively. Now, for $F_0$ prediction, we train another CycleGAN architecture, where $y' \epsilon R^N$ is the cepstral features of converted normal, which is extracted from previously trained CycleGAN for MCC mapping, and $z \epsilon R^1$ is the $F_0$ of original normal speech.

\section{Proposed CinC-GAN}
\label{sec:cinc}

\textbf{Problem formulation:} The conventional formulation for WHSP2SPCH conversion is $y' = f(x) + n$ for cepstral feature mapping, where $x$ is whisper speech features, $f$ is the mapping function, and $n$ is the additive noise. Now, for $F_0$ prediction, we formulate the problem as $z = g(y') + n'$, which implies that $z = g(f(x) + n) + n'$, where $g$ is the mapping function, and $n'$ another additive noise.

Given this problem, we observed that due to the use of two differently trained mapping functions, for $F_0$ prediction, significant non-linear noise is being added. Hence, for effective $F_0$ prediction and to suppress this noise, we need some sophisticated mapping function, which can be trained simultaneously, and somehow it can also directly rely on input instead of only $f(x)$.

\textbf{Proposed solution:} In this paper, we propose a different training method, namely, Cycle-in-Cycle GAN (CinC-GAN), which is an advanced version of CycleGAN, for WHSP2SPCH conversion. In CycleGAN, we use one model for acoustic feature mapping, and second for $F_0$ prediction, where both of them are separately trained (i.e., sequential training). However, in CinC-GAN, we use inner cycle for acoustic feature mapping, and outer cycle for $F_0$ prediction, where outer cycle relies on cepstral features of converted normal speech, and input whisper speech as well (i.e., joint training). This way, we are able to achieve our goal, and suppress the effect of extra noise.

In summary, we propose a Cycle-in-Cycle GAN as shown in Fig. \ref{fig:cincgan}. In this approach, we adopt two coupled CycleGANs to learn the mapping for $X$ to $Y$ and $Y$ to $Z$, respectively. In addition, non-parallel dataset $x \epsilon X$, $y \epsilon Y$, and $z \epsilon Z$ is used for training, where $X$ and $Y$ are set of cepstral features of whisper and normal speech, respectively, and $Z$ is set of $F_0$ extracted from the normal speech. Detailed description on feature extraction is given in Section \ref{sec:exprmnts}.

\subsection{Acoustic feature mapping}

\begin{figure}
    \centering
    \includegraphics[width=\linewidth]{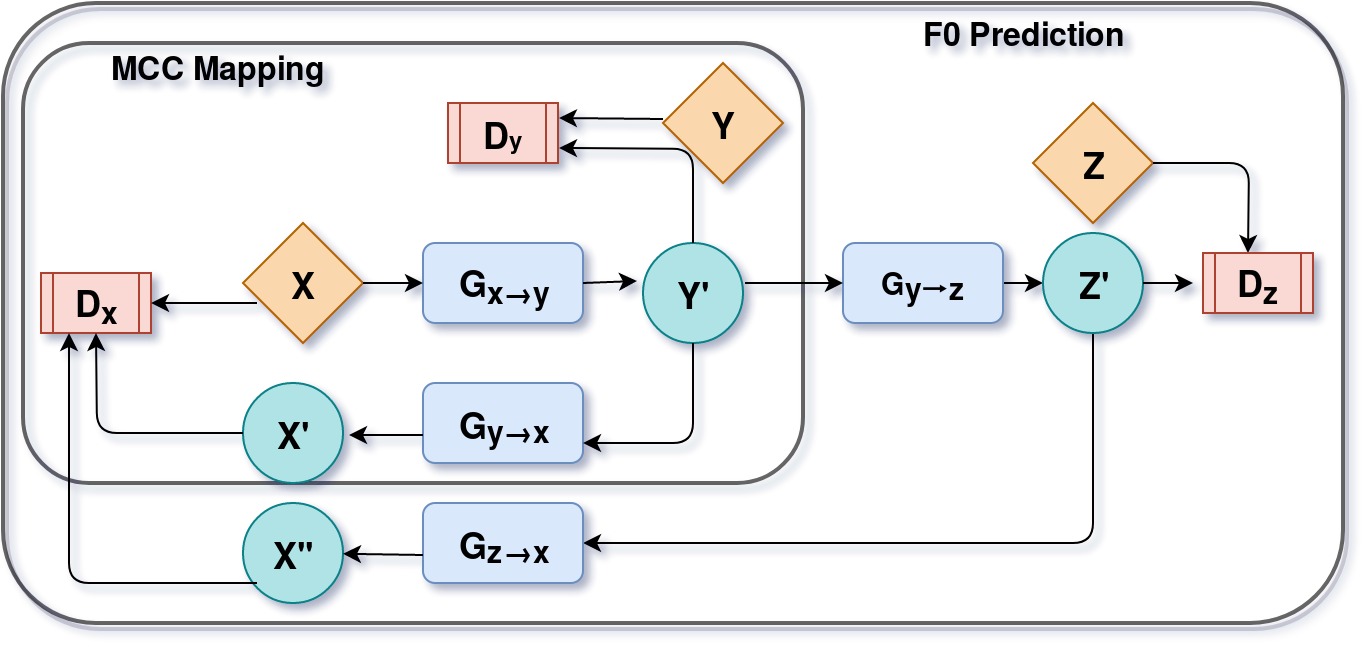}
    \caption{Proposed Cycle-in-Cycle GAN. After \cite{yuan2018unsupervised}.}
    \label{fig:cincgan}
\end{figure}

The inner cycle in Fig. (\ref{fig:cincgan}) maps cepstral features of whisper ($X$) to normal speech ($Y$). We use two generators, $G_{X \rightarrow Y}$ and $G_{Y \rightarrow X}$, where $G_{X \rightarrow Y}$ maps $x$ to $Y$ and $G_{Y \rightarrow X}$ maps $y$ to $X$. The discriminators $D_X$ and $D_Y$ confirms whether generated distribution is from $X$ and $Y$ or not, respectively. Here, we use adversarial loss, cycle-consistency loss, and identity loss. Adversarial loss is defined as:

% The generator $G_{X \rightarrow Y}$ learns to generate normal speech features, and discriminator $D_Y$ learns to distinguish between generated and original normal speeches. Therefore, it is used to make generated features more similar to the original normal speech features.

\begin{equation}
\begin{split}
\mathcal{L}_{adv_1}(G_{X \rightarrow Y},& D_Y) =  \mathbb{E}_{y \sim P_{Y}(y)}[(D_Y(y) - 1)^2] \\ &+ \mathbb{E}_{x \sim P_X(x)}[(D_Y(G_{X \rightarrow Y}(x)))^2].
\end{split}
\label{eq:cycle_adv1}
\end{equation}

To map the two different distributions (i.e., normal and whisper speech), we add generator $G_{Y \rightarrow X}$ to map normal-to-whisper speech features. In addition, we use discriminator, $D_X$ to distinguish between real and generated whisper speech. Therefore, we also use single cycle-consistency loss: i.e.,

\begin{equation}
\begin{split}
\mathcal{L}_{cyc_1}(G_{X \rightarrow Y}) =  \mathbb{E}_{x \sim P_X(x)} [\| G_{Y \rightarrow X}(G_{X \rightarrow Y}(x)) - x\|_1 ].
\end{split}
\label{eq:cycle_cyc1}
\end{equation}

In addition, we use identity loss to preserve the linguistic content, i.e.,

\begin{equation}
\begin{split}
\mathcal{L}_{id_1}(G_{X \rightarrow Y}, G_{Y \rightarrow X}) &=  \mathbb{E}_{x \sim P_X(x)} [\| G_{Y \rightarrow X}(x) - x\|_1 ] \\ &+ \mathbb{E}_{y \sim P_Y(y)} [\| G_{X \rightarrow Y}(y) - y\|_1].
\end{split}
\label{eq:cycle_idnt1}
\end{equation}

\subsection{$F_0$ Prediction}

After mapping the cepstral features of whisper-to-normal speech, we focus on $F_0$ prediction task. Previous methods tries to predict $F_0$ from the cepstral features of converted normal speech using CycleGAN, which is trained separately (i.e., sequential training). However, in this paper, we propose to predict $F_0$ from the cepstral features of converted normal speech simultaneously via joint training.

 We use the generator $G_{Y \rightarrow Z}$ to predict $F_0$ from the converted normal speech ($G_{X \rightarrow Y}(x)$) and $G_{Z \rightarrow X}$ is used to map the predicted $F_0$ to whisper speech instead of normal speech. This way, we are able to remove the non-linear noise by including the effect of original whisper speech and joint training methodology. In addition, we use discriminator $D_Z$ to make generated $F_0$ just like original $F_0$. However, to add the effect of whisper speech, we used fourth generator to generate whisper speech features from the predicted $F_0$ instead to converted normal speech features. Here, we adapt only two losses, adversarial loss, and cycle-consistency loss, i.e.,
 
% In short, the generator $G_{Y \rightarrow Z}$ to predict $F_0$ from the converted normal speech ($G_{X \rightarrow Y}(x)$) will be trained parallel with the generator $G_{Y \rightarrow Z}$. This way, we are able to remove the non-linear noise via joint training and including the effect of original whisper speech. This way, we are able to remove the non-linear noise.

\begin{equation}
\begin{split}
\mathcal{L}_{adv_2}(G_{Y \rightarrow Z},& D_Z) =  \mathbb{E}_{z \sim P_{Z}(z)}[(D_z(z) - 1)^2] \\ &+ \mathbb{E}_{y \sim P_Y(y)}[(D_Z(G_{Y \rightarrow Z}(y)))^2].
\end{split}
\label{eq:cycle_adv2}
\end{equation}

\vspace{-0.3cm}

\begin{equation}
\begin{split}
&\mathcal{L}_{cyc_2}(G_{Y \rightarrow Z}) \\ &=  \mathbb{E}_{y \sim P_Y(y)} [\| G_{Z \rightarrow Y}(G_{Y \rightarrow Z}(y)) - y\|_1 ].
\end{split}
\label{eq:cycle_cyc2}
\end{equation}

\vspace{0.3cm}

Moreover, we add combine loss through a third discriminator $D_{X}$. This discriminator confirms the output of two generators ($G_{Y \rightarrow X}$, $G_{Z \rightarrow X}$) is from original distribution of $X$ or not. This way both (inner and outer) cycles stay connected with common measure of reconstruction.

\begin{equation}
\begin{split}
& \mathcal{L}_{adv_3}(G_{Y \rightarrow X}, G_{Z \rightarrow X}) \\ &=  \mathbb{E}_{x \sim P_{X}(x)}[(D_X(x) - 1)^2] \\ &+ \mathbb{E}_{y \sim P_Y(y)}[(D_X(G_{Y \rightarrow X}(G_{X \rightarrow Y}(x))))^2] \\ &+ \mathbb{E}_{z \sim P_Z(z)}[(D_X(G_{Z \rightarrow X}(G_{Y \rightarrow Z}(G_{X \rightarrow Y}(x)))))^2].
\end{split}
\label{eq:cycle_adv3}
\end{equation}

\vspace{0.1cm}

\subsection{Overall Objective of the Proposed Method}

In summary, we train both the cycles simultaneously. And we optimize all the generators, and discriminators according to the following rules:

\begin{equation}
\begin{split}
&\mathcal{L}_{full} =  \mathcal{L}_{adv_1} + \lambda_1*\mathcal{L}_{cyc_1} + \lambda_2*\mathcal{L}_{id_1} + \\ & \lambda_3*\mathcal{L}_{adv_2} + \lambda_4*\mathcal{L}_{cyc_2} + \lambda_5*\mathcal{L}_{adv_3},
\end{split}
\label{eq:cycle_full}
\end{equation}

where  $\lambda_1$, $\lambda_2$, $\lambda_3$, $\lambda_4$, and $\lambda_5$ are the hyperparameters associated with different loss functions. These parameters defines relative importance of each losses w.r.t. the other losses. Here, $\lambda_1 = 10$, $\lambda_2 = 5$, $\lambda_3 = 10$, $\lambda_4 = 1$, and $\lambda_5 = 1$ are used \textit{empirically} in all of our experiments (because this choice of hyperparameters shows stable and accurate training). And these hyperparameter values work for any conversion pairs.

\section{Experimental Results}
\label{sec:exprmnts}

\subsection{Dataset and Feature Extraction}
In WHSP2SPCH conversion, we have used Whispered TIMIT (wTIMIT) database \cite{lim2011computational}. In both the approaches, i.e., speaker-specific and gender-specific, we have done non-parallel training. We have done speaker-specific task on four different speakers, specifically two female and two male speakers. Particularly, for each speaker, $34$ minutes of training data and $2.32$ minutes of testing data was used. In each gender-specific task, we have used four speakers, and particularly, in each training, $136$ minutes of training data was used. In gender-specific task, we test it on four seen and two unseen speakers, and test data for each speaker is $13.92$ minutes. We extract the $F_0$ and MCC (Mel Cepstral Coefficient) features from whisper and normal speech using AHOCODER \cite{erro2011improved}. In feature extraction, we have used 25 ms window size, and 5 ms frame shift \cite{erro2011improved}.

\subsection{Architecture Details}
\label{architecture}

Generators $G_{X \rightarrow Y}$, $G_{Y \rightarrow X}$ and $G_{Y \rightarrow Z}$ follow the same configuration, for both the architectures. In $G_{X \rightarrow Y}$ and $G_{Y \rightarrow X}$, contain 40, 512, and 40 neurons in input layer, hidden layers, and output layer, respectively. Generator $G_{Y \rightarrow Z}$ contains 40, 512, and 1 neurons in input layer, hidden layers and output layer, respectively. $G_{Z \rightarrow Y}$ has the 1, 512, and 40 neurons in input layer, hidden layers, and output layer, respectively. All layers are followed by Rectified Linear Unit (ReLU) activation function. All discriminators follow the same configuration for both the architecture. $D_X$, $D_Y$, and $D_Y'$ have the 40, 512, and 1 neurons in the input layer, hidden layers, and output layer, respectively. $D_Z$ has the 1, 512, and 1 neurons in the input layer, hidden layers, and output layer, respectively. In all discriminators, input layer and all hidden layers are followed by \textit{ReLU} activation function and output layer followed by \textit{sigmoid} activation function. Both the architectures are trained for 100 epochs, and learning rate was set to $0.0001$. Source code is provided at \href{https://github.com/Maitreyapatel/speech-conversion-between-different-modalities}{https://github.com/Maitreyapatel/speech-conversion-between-different-modalities}.  

% \footnote{We highly encourage to refer this link for code and samples: \url{https://tinyurl.com/y4s3yg3e}}

\vspace{7mm}

\subsection{Objective Evaluation}
\label{sec:objevl}

We have applied Mel Cepstral Distortion (MCD), and Root Mean Square Error (RMSE) of log($F_0$)-based objective measures to analyze the effectiveness of the WHSP2SPCH conversion systems \cite{toda2007voice}. MCD is the distance between the converted and the reference cepstral features, a system that is having lesser MCD is considered as a better system. Lesser the RMSE of $log(F_0)$, better the system is.

% To measure the RMSE of $\log(F_0)$, the actual reference speech, and the converted speech signals, are time-aligned using the Dynamic Time Warping (DTW) algorithm. These DTW aligned pairs will generate voiced-voiced (VV), voiced-unvoiced (VU), unvoiced-voiced (UV), and unvoiced-unvoiced (UU) pairs. Here, we consider only VV pairs for computing the RMSE of the $\log(F_0)$ \cite{wu2010text}. Lesser the RMSE of $\log(F_0)$, better the system is.

\begin{table}[t]
\caption{MCD analysis of the different WHSP2SPCH systems for speaker-specific task. Here, \% in the bracket indicates the relative reduction in the MCD w.r.t the baseline}
\label{tab:my-table1}
\resizebox{\linewidth}{!}{
\begin{tabular}{|l|c|c|c|c|}
\hline
\backslashbox[40mm]{\textbf{Method}}{\textbf{Speaker}} & \textbf{\begin{tabular}[c]{@{}c@{}}F1\\ (US\_102)\end{tabular}} & \textbf{\begin{tabular}[c]{@{}c@{}}M1\\ (US\_103)\end{tabular}} & \textbf{\begin{tabular}[c]{@{}c@{}}F2\\ (US\_104)\end{tabular}} & \textbf{\begin{tabular}[c]{@{}c@{}}M2\\ (US\_106)\end{tabular}} \\ \hline
\textbf{CycleGAN (Baseline)} & 6.76 & 6.36 & 6.1 & 5.97 \\ \hline
\textbf{CinC-GAN} & \begin{tabular}[c]{@{}c@{}}6.73\\ (0.4\%)\end{tabular} & \begin{tabular}[c]{@{}c@{}}6.42\\ (-0.94\%)\end{tabular} & \begin{tabular}[c]{@{}c@{}}6.11\\ (-0.1\%)\end{tabular} & \begin{tabular}[c]{@{}c@{}}5.86\\ (2\%)\end{tabular} \\ \hline
% \textbf{CinC-GAN + Regularizer} & \begin{tabular}[c]{@{}c@{}}6.74\\ (0.29\%)\end{tabular} & \begin{tabular}[c]{@{}c@{}}6.4\\ (-0.63\%)\end{tabular} & \begin{tabular}[c]{@{}c@{}}6.09\\ (0.16\%)\end{tabular} & \begin{tabular}[c]{@{}c@{}}5.96\\ (0.17\%)\end{tabular} \\ \hline
\end{tabular}
}
\end{table}

% \multicolumn{1}{|c|}{\textbf{Method/Speaker}}

\begin{table}[t]
\caption{MCD analysis of the different WHSP2SPCH systems for gender-specific task. Here, \% in the bracket indicates the relative reduction in the MCD w.r.t the baseline}
\label{tab:my-table3}
\resizebox{\linewidth}{!}{
\begin{tabular}{|l|c|c|c|c|}
\hline
\backslashbox[40mm]{\textbf{Method}}{\textbf{Speaker}} & \textbf{F-Seen} & \textbf{M-Seen} & \textbf{F-Unseen} & \textbf{M-Unseen} \\ \hline
\textbf{CycleGAN (Baseline)} & 6.69 & 6.28 & 6.77 & 6.83 \\ \hline
\textbf{CinC-GAN} & \begin{tabular}[c]{@{}c@{}}6.66\\ (0.45\%)\end{tabular} & \begin{tabular}[c]{@{}c@{}}6.29\\ (-0.16\%)\end{tabular} & \begin{tabular}[c]{@{}c@{}}6.92\\ (-2.2\%)\end{tabular} & \begin{tabular}[c]{@{}c@{}}6.9\\ (-1.0\%)\end{tabular} \\ \hline
\end{tabular}
}
\end{table} 

\begin{table}[t]
\caption{RMSE-based objective analysis of $\log(F_0)$  for speaker-specific task. Here, \% in the bracket indicates a relative reduction in the RMSE w.r.t the baseline}
\label{tab:my-table2}
\resizebox{\linewidth}{!}{
\begin{tabular}{|l|c|c|c|c|}
\hline
\backslashbox[40mm]{\textbf{Method}}{\textbf{Speaker}} & \textbf{\begin{tabular}[c]{@{}c@{}}F1\\ (US\_102)\end{tabular}} & \textbf{\begin{tabular}[c]{@{}c@{}}M1\\ (US\_103)\end{tabular}} & \textbf{\begin{tabular}[c]{@{}c@{}}F2\\ (US\_104)\end{tabular}} & \textbf{\begin{tabular}[c]{@{}c@{}}M2\\ (US\_106)\end{tabular}} \\ \hline
\textbf{CycleGAN (Baseline)} & 7.19 & 5.7 & 3.88 & 6.49 \\ \hline
\textbf{CinC-GAN} & \begin{tabular}[c]{@{}c@{}}5.65\\ (\textbf{21.4\%})\end{tabular} & \begin{tabular}[c]{@{}c@{}}4.6\\ (\textbf{19.3\%})\end{tabular} & \begin{tabular}[c]{@{}c@{}}2.77\\ (\textbf{28.5\%})\end{tabular} & \begin{tabular}[c]{@{}c@{}}3.25\\ (\textbf{49.9\%})\end{tabular} \\ \hline
% \textbf{CinC-GAN + Regularizer} & \begin{tabular}[c]{@{}c@{}}6.4\\ (11\%)\end{tabular} & \begin{tabular}[c]{@{}c@{}}4\\ (22.8\%)\end{tabular} & \begin{tabular}[c]{@{}c@{}}3.85\\ (22.8\%)\end{tabular} & \begin{tabular}[c]{@{}c@{}}5.9\\ (9.09\%)\end{tabular} \\ \hline
\end{tabular}
}
\end{table}

\begin{table}[t]
\caption{RMSE-based objective analysis of $\log(F_0)$ for gender-specific task. Here, \% in the bracket indicates a relative reduction in the RMSE w.r.t the baseline}
\label{tab:my-table4}
\resizebox{\linewidth}{!}{
\begin{tabular}{|l|c|c|c|c|}
\hline
\backslashbox[40mm]{\textbf{Method}}{\textbf{Speaker}} & \textbf{F-Seen} & \textbf{M-Seen} & \textbf{F-Unseen} & \textbf{M-Unseen} \\ \hline
\textbf{CycleGAN (Baseline)} & 18.2 & 38.9 & 25.6 & 28.3 \\ \hline
\textbf{CinC-GAN} & \begin{tabular}[c]{@{}c@{}}6.81\\ (\textbf{62.3\%})\end{tabular} & \begin{tabular}[c]{@{}c@{}}3.16\\ (\textbf{91.9\%})\end{tabular} & \begin{tabular}[c]{@{}c@{}}3.14\\ (\textbf{87.7\%})\end{tabular} & \begin{tabular}[c]{@{}c@{}}3.8\\ (\textbf{86.6\%})\end{tabular} \\ \hline
\end{tabular}
}
\end{table}

The effectiveness of CinC-GAN can be clearly seen for the WHSP2SPCH conversion system in objective results. Analysis of both the architectures is done using 2 different approaches 1) speaker-specific in which is model is trained an tested only on single speaker and 2) gender-specific in which model is trained for specific number of speakers and tested on seen as well as out of the box speaker (unseen speaker). As shown in Table \ref{tab:my-table1}, it can be observed that CinC-GAN performs comparatively to CycleGAN in terms of MCD. However, CinC-GAN outperforms CycleGAN in terms of RMSE $\log(F_0)$ for all the speakers (as shown in Table \ref{tab:my-table2}). CinC-GAN gets on an average 29.8\% relative reduction in case of speaker-specific, compared to the CycleGAN in $F_0$-$RMSE$. Moreover, Table \ref{tab:kl_speaker} shows the Kullback-Leibler Divergence (KLD) and Jensen-Shannon Divergence (JSD) between predicted $F_0$ and original $F_0$ for speaker-specific task. Here, we can observed that CinC-GAN outperforms CycleGAN. Therefore, this analysis further strengthens our results.

\begin{table}[H]
\caption{Results of KL-JSD for Speaker-specific task.}
\label{tab:kl_speaker}
\centering
\resizebox{.75\linewidth}{!}{
\begin{tabular}{|c|c|c|c|c|}
\hline
\backslashbox[41mm]{\textbf{Speaker}}{\textbf{Method}} & \multicolumn{2}{c|}{\textbf{CinC-GAN}} & \multicolumn{2}{c|}{\textbf{CycleGAN}} \\ \cline{2-5} 
                 & \textbf{KL}   & \textbf{JSD}  & \textbf{KL} & \textbf{JSD} \\ \hline
\textbf{US\_102} & 5.11 & 4.95 & 29.23 & 4.76        \\ \hline
\textbf{US\_103} & 5.27 & 5.83 & 0.03 & 7.22         \\ \hline
\textbf{US\_104} & 7.29 & 4.46 & 20.94 & 7.74        \\ \hline
\textbf{US\_106} & 2.37 & 3.59 & 4.85 & 1.27         \\ \hline
\textbf{Average} & \textbf{7.51} & \textbf{4.71} & 13.76 & 5.25         \\ \hline
\end{tabular}
}
\end{table}

We further extend our experiment, and perform objective evaluation for gender-specific task. For this, we trained two CinC-GAN, first on 4 female speakers, and second on 4 male speakers. We tested both of them on seen speaker and unseen utterances, and unseen speaker, as well. As shown in Table \ref{tab:my-table3}, in terms of MCD, CycleGAN and CinC-GAN performs similarly. However, in terms of $F_0$-$RMSE$ CinC-GAN outperforms CycleGAN by on an average 82.1\%, as shown in Table \ref{tab:my-table4}. We observed that the CycleGAN is not able to predict $F_0$ effectively on combined dataset, whereas CinC-GAN works quite efficiently in every scenarios even on unseen speaker and unseen utterances.

% We tested both of them on seen speaker and unseen utterances, and unseen speaker and unseen utterances, as well.

% We conducted speaker (trained and tested on single speaker only) and gender (trained on 4 speakers with same gender and tested on these and out of box same gender speakers) specific analysis.

\subsection{Subjective Evaluation}
\label{sec:subevl}
\vspace{-0.5cm}
\begin{figure}[!h]
    \centering
    \includegraphics[width=\linewidth]{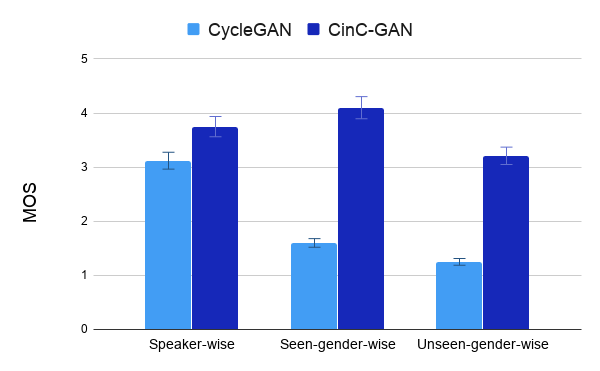}
    \caption{MOS score analysis for speaker-specific and gender-specific task (i.e., seen-unseen) with $95\%$ confidence interval.}
    \label{fig:mos}
\end{figure}

For subjective test analysis, Mean Opinion Score (MOS) has been taken to measure the naturalness of the converted speech. Total 28 subjects (7 females and 21 males between 18 to 30 years of age and with no known hearing impairments) took part in the subjective test. Here, we randomly played utterances from both the systems. In the MOS test, subjects were asked to rate the played utterances on the scale of $1$-$5$, where $1$ indicates completely whisper speech, and $5$ means completely converted in normal speech. We can observe that the CinC-GAN has almost $20.2$\% more naturalness in case of speaker-specific task. From Fig. \ref{fig:mos}, we can observe that CinC-GAN significantly outperforms CycleGAN for seen and unseen (out-of-the-box) speakers, respectively, on gender-specific task. In addition, in this case, CycleGAN fails measurably and produces whisper speech even for seen and unseen speakers, which can be observed in MOS plot shown in Fig. \ref{fig:mos}. However, CinC-GAN maintains its performance for seen and unseen speakers. CinC-GAN is able to score $MOS\geq3$ for gender-specific task for unseen speaker as well. Therefore, CinC-GAN leads to the possibility of few-shot learning for WHSP2SPCH for the first time in literature. 

% \begin{table}[h]
% \caption{KL-JSD for Speaker specific}
% \label{tab:kl_speaker}
% \resizebox{\linewidth}{!}{
% \begin{tabular}{|l|c|c|c|c|c|}
% \hline
% Speakers/Architecture & CinCGAN-KL & CycleGAN-KL & CinCGAN-JSD & CycleGAN-JSD \\ \hline
% US\_102 & 15.11 & 29.23 & 4.95 & 4.76  \\ \hline
% US\_103 & 5.27  & 0.03  & 5.83 & 7.22  \\ \hline
% US\_104 & 7.29  & 20.94 & 4.46 & 7.74  \\ \hline
% US\_106 & 2.37  & 4.85  & 3.59 & 1.27  \\ \hline
% Average & \textbf{7.51}  & 13.76 & \textbf{4.71} & 5.25  \\ \hline
% \end{tabular}
% }
% \end{table}

% Results of Gender_specific (Average values of all the speakers - seen and unseen):

% Male - 
%      CinCGAN   CycleGAN
% KL    10.96     17.44
% JSD   2.10      5.33

% Female -
%      CinCGAN   CycleGAN
% KL    27.53     17.77
% JSD   5.22      5.66

% From the converted cepstral features and predicted $F_0$, we synthesis speech using AHODECODER. We compared the original speech and speech which is reconstructed from the features of that same file and we observed that, AHODECODER adds quite noise during speech synthesis, which can leads to reduce in MOS.

% \vspace{-2mm}

\section{Summary and Conclusion}
\label{sec:conclusion}
% \vspace{-2mm}
In this paper, we proposed the CinC-GAN to increase the effectiveness of $F_0$ prediction without affecting accuracy of MCC mapping. Baseline (i.e., CycleGAN) uses sequential training, which adds non-linear noise in $F_0$ prediction. However, CinC-GAN adopts joint training methodology to decrease this noise. Objective and subjective results show superiority of CinC-GAN over the baseline. In addition, CycleGAN fails in WHSP2SPCH conversion for gender-specific task. However, CinC-GAN maintains its result even for out-of-the-box speaker. This shows the potential of CinC-GAN for few-shot WHSP2SPCH conversion. In future, we plan to extend our study on zero-shot and one-shot WHSP2SPCH conversion.

\bibliographystyle{IEEEbib}
\bibliography{refs}
% ...
% ...
% \printbibliography

\end{document}